\documentclass [aps,pre,onecolumn,preprint,superscriptaddress] {revtex4-1}
\usepackage{graphicx}
\usepackage{subfigure}
\usepackage{amsmath}
\usepackage{xcolor}
\usepackage{hyperref}

\begin{document}
\title{Electrowetting-Controlled Dropwise Condensation with Patterned Electrodes: Physical Principles, Modeling, and Application Perspectives for Fog Harvesting and Enhanced Heat Transfer}

\author{Harmen Hoek}
\affiliation{Physics of Complex Fluids, MESA+ Institute for Nanotechnology, University of Twente, P.O. Box 217, 7500 AE Enschede, The Netherlands}

\author{Ranabir Dey}
\affiliation{Dynamics of Complex Fluids, Max Planck Institute for Dynamics and Self‐-organization, Am Fassberg 17, Goettingen, 37077 Germany}

\author{Frieder Mugele}
\affiliation{Physics of Complex Fluids, MESA+ Institute for Nanotechnology, University of Twente, P.O. Box 217, 7500 AE Enschede, The Netherlands}

%\myworries{}

\begin{abstract}
Patterning the wettability of solid surfaces is a successful strategy to control the dropwise condensation of vapor onto partially wetting solid surfaces. We followed the condensation of water vapor onto electrowetting-functionalized surfaces with structured co-planar electrodes. A detailed analysis of the experimental distribution of millions of drops reveals that despite the presence of contact angle hysteresis and the occurrence of random drop coalescence events, the preferential drop position closely follows the evolution of the local minima of the numerically calculated drop size-dependent electrostatic energy landscape in two dimensions. Even subtle transitions between competing preferred locations are properly reproduced by the model. Based on this quantitative understanding of the condensation patterns, we discuss a series of important follow-up steps that need to be taken to demonstrate a reliable performance gain in various applications focusing in particular on enhanced heat transfer.
\end{abstract}
\maketitle

\section{Introduction}
The condensation of water vapor onto solid surfaces is integral to many natural processes including dew formation\cite{Beysens1995} and fog harvesting by animals, like the Namib Desert Beetle\cite{Parker2001,Park2016} and Litoria caerulea-, a green tree frog in Australia,\cite{Tracy2011} and plants, such as the Namib desert plant.\cite{Malik2014} Water condensation is also intrinsic to various technological applications like fog harvesting,\cite{Milani2011,Lee2012} seawater desalination,\cite{Khawaji2008} and heat exchangers for power generation\cite{Beer2007} and refrigeration.\cite{Barbosa2012,Kim2002} In all cases, efficient condensation and removal (or `collection') of the condensed liquid is essential. The entire process consists of a series of steps, namely the nucleation of liquid on an initially dry solid surface, the subsequent growth of the liquid phase in form of a film or droplets, and finally the removal of the latter. At first glance, hydrophilic surfaces may seem the most natural choice to promote condensation. Yet, it has been known for decades that plain hydrophilic surfaces are actually not the best choice because they promote the formation of condensed liquid film (for a review: see \cite{Rose2002}). Compared to films, drops are much easier to manipulate and transport in desired directions by suitable topographical and chemical patterns on the surface. Moreover, particularly in heat transfer, thick films of condensed liquid form a barrier of poor thermal conductivity that prevents direct contact of the vapor with the cooled surface of the condenser and thus reduces the overall heat transfer. Hence, it is usually advantageous to use partially wetting solid surfaces where condensing vapor forms discrete drops that leave parts of the solid surface in direct contact with the to-be-condensed vapor. As these discrete drops are removed, they expose even more bare surface again and thereby free space for a subsequent generation of condensing drops. Like in case of biological or technological fog harvesting surfaces, efficient removal of the condensate drops is therefore essential for the overall performance of the system. 

Throughout recent years, various efforts have been made to optimize dropwise condensation and the subsequent removal of drops using suitable topographical and chemical surface patterns.\cite{Hou2015,Mondal2015,Ghosh2014,Narhe2004,Boreyko2009,Miljkovic2012,Miljkovic2013b,Weisensee2017,Anand2012,Tsuchiya2017} Such patterns generate an energy landscape in which condensing drops initially form at either random locations or at preferred hydrophilic nucleation sites. As drops grow with time, they experience the imprinted gradients in wettability, hit geometric boundaries, and coalesce with other drops. In each of these situations, the original configuration of the drop typically becomes unstable and the drop moves towards a location of lower energy. Examples of such surface patterns include surfaces with alternating hydrophobic and hydrophilic stripes,\cite{Hou2015,Mondal2015,Ghosh2014} surfaces with conical geometries,\cite{Park2016} superhydrophobic surfaces with grooves\cite{Narhe2004} or nanostructures,\cite{Boreyko2009,Miljkovic2012,Miljkovic2013b} as well as liquid-infused surfaces.\cite{Weisensee2017,Anand2012,Tsuchiya2017} The resulting drop displacements are either driven entirely by capillary and wetting forces or they may be assisted by gravity in case of vertically oriented condenser surfaces. In all cases, drops only move once the driving forces are strong enough to overcome the pinning due to microscopic heterogeneities.\cite{DeRuiter2015} The latter are usually quantified by specifying the contact angle hysteresis $\Delta \cos\theta=\cos\theta_r-\cos\theta_a$, where, $\theta_r$ and $\theta_a$ are the receding and advancing contact angles. This explains the interest in surfaces with low contact angle hysteresis such as superhydrophobic and liquid-infused surfaces for heat transfer applications with dropwise condensation. 

The approaches described above all rely on passive wettability patterns imprinted onto the solid surface upon fabrication. In contrast, electrowetting (EW) allows for active tuning of the wettability and controlled transport of drops of conductive liquids such as water on partially wetting hydrophobic surfaces.\cite{Pollack2000,Cho2003,Mugele2005,MugeleBook} While generically used in combination with a wire that is immersed directly into the liquid, capacitive coupling between the drop(s) and suitably structured co-planar electrodes on the substrate that are covered by a thin hydrophobic polymer layer allow for similarly efficient control of the wettability locally above the activated electrodes.\cite{Yi2006,MugeleBook} By patterning the electrodes, wettability patterns such as simple traps for drops can be generated and switched on and off at will.\cite{Mannetje2013} Drops that were large compared to the width of a gap between two electrodes, preferentially aligned on the center of the gap. As usual in EW, this minimum of the electrostatic energy $E_{el} =-C_{tot}U^2/2$ corresponds to the maximum of the total capacitance between the drop and the electrodes. In this manner, 't Mannetje et al. \cite{TMannetje2014} demonstrated controlled capture, release, and steering of rolling drops on an inclined plane. Later de Ruiter et al.\cite{DeRuiter2014} extended the same principle for drops in microfluidic two phase flow systems for a range of electrode geometries and applied a simple analytical model to calculate the electrical holding force based on the geometric overlap of the trapped drop and the activated electrodes. The idea of manipulating condensing drops by EW was first explored by Kim and Kaviany.\cite{Kim2007} Baratian et al.\cite{Baratian2018} later combined these ideas to study for the first time directly the condensation of water vapor onto EW-functionalized surfaces. For the specific case of parallel interdigitated electrodes aligned along the direction of gravity, they found that the condensation pattern is governed by an electrostatic energy landscape that depends on the size of the condensing drops. While the initial condensation occurred at random locations, subsequent growth by further condensation and EW-induced coalescence lead to alignment of the drops along the edges of the electrodes. Later, once their diameter became comparable to the width of the electrodes, the drops accumulated at the centers of the gaps between adjacent electrodes. Analyzing the distribution of drop sizes and locations, they showed that the drops decorate the drop size-dependent minima of the (one-dimensional) electrostatic energy landscape perpendicular to the electrodes. EW-induced coalescence events lead to faster drop growth. In combination with the reduced contact angle hysteresis in EW with AC voltage\cite{Li2008} drop shedding occurs on average for smaller drops, as compared to the reference case without EW.\cite{Baratian2018} According to classical observations in dropwise condensation, such a reduction of the critical shedding radius is accompanied by enhanced heat transfer.\cite{Rose2002}

A series of follow-up studies confirmed these basic original observations regarding the evolution of the drop distribution for straight interdigitated electrodes.\cite{Yan2018,Yan2019a,Wikramanayake2019,Wikramanayake2020a,Hognadottir2020} Experiments with slightly more complex electrode geometries with zigzag-shaped edges resulted in preferential alignment of the drops not only perpendicular but also along the direction of the electrodes, in qualitative agreement with expectations.\cite{Dey2018} That study also indirectly inferred an increased heat transfer from the volume of shedded drops as extracted from video microscopy images. Overall, the experiments suggest that it should be possible to optimize the performance of EW-controlled condensation in heat transfer and other applications by systematically varying electrode geometries and/or excitation patterns. Since experimental brute force optimization of electrode shapes would be very time consuming and costly, it is essential then to extend the existing electrostatic models to arbitrary electrode geometries, and to demonstrate their performance in capturing the complex evolution of drop distribution patterns to enable electrode optimization \emph{in silico} prior to experimental testing. 

The purpose of the present work is therefore twofold: the core of the work consists of a detailed comparison of the distribution of approximately 87 million drops with sizes between 4.3 and 2000 $\mu$m extracted using image analysis with the predictions of a numerical model based on the drop size-dependent minimization of the electrostatic energy. Experiments and calculations are carried out for the specific case of interdigitated electrodes with zigzag shaped edges of variable length. The comparison reveals an impressive degree of agreement and correctly reproduces a series of subsequent transitions of preferred drop positions as a function of size. The previously proposed simple analytical model by 't Mannetje et al.\cite{Mannetje2013} reproduces the qualitative behavior but underestimates electrostatic energies and forces for small drops. Following the discussion of these results, we evaluate the present status of the field and discuss aspects that we consider essential for the development of EW-controlled condensation from a physical phenomenon towards a technologically relevant application.

\section{Methods}

\subsection{Experimental Aspects}
\begin{figure}[ht]
\centering
  \includegraphics[width=1\linewidth]{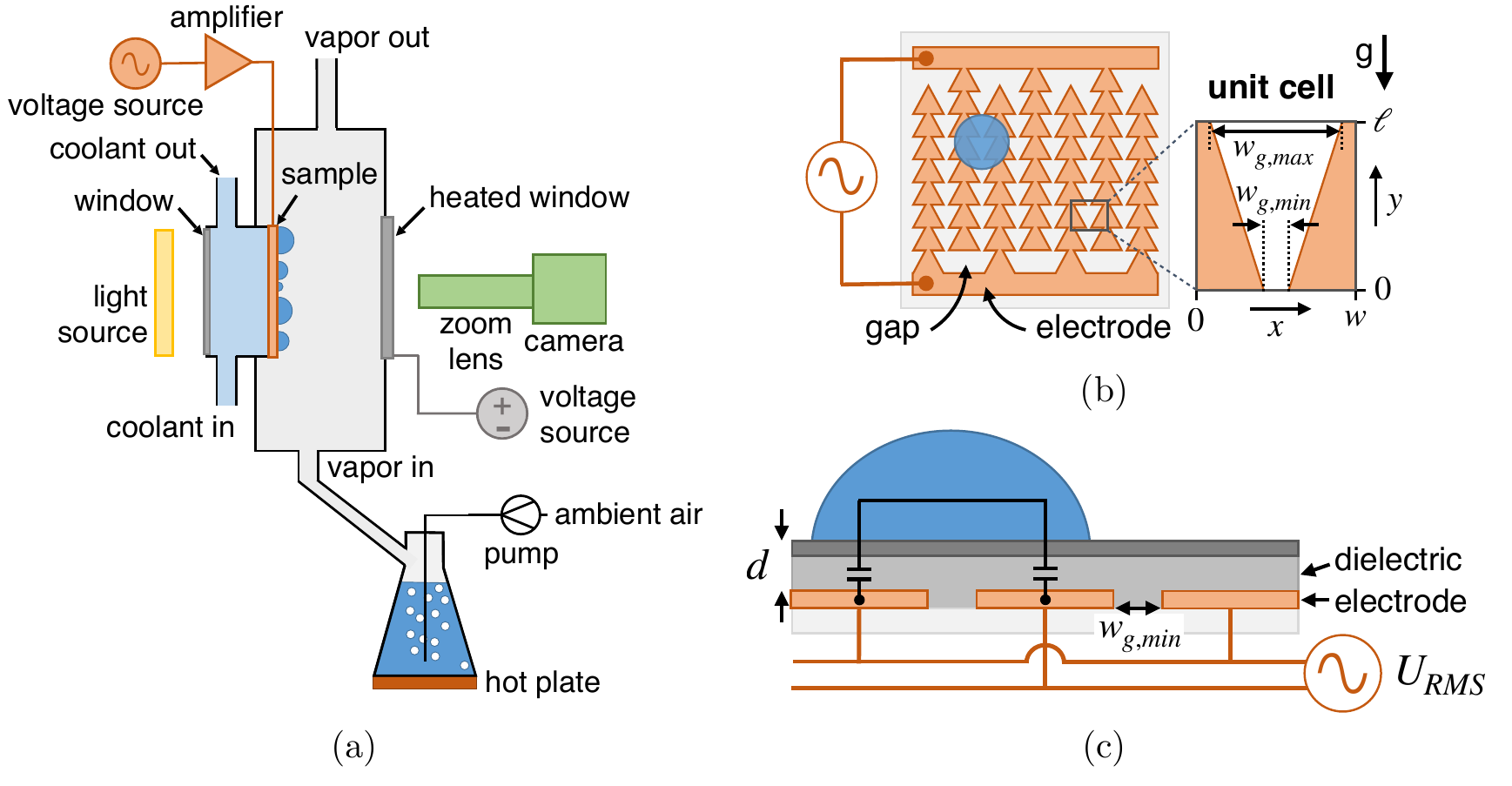}
\caption{Experimental setup (not to scale). (a) Schematic of vapor generator, condensation chamber, cooled sample stage, and optical setup (not to scale). (b) Top view of vertically oriented sample with zoomed view of unit cell of electrode pattern. (c) Cross-sectional view of a condensed drop on the substrate.}
\label{fig:experimentalsetup}
\end{figure}

The present condensation experiments were performed in the same homemade experimental setup (Figure \ref{fig:experimentalsetup}a) that was used in our previous studies.\cite{Baratian2018,Dey2018} The setup consists of a condensation chamber with two inlets at the bottom and an outlet through a fine grid of holes for vapor at the top side. The transparent sample is mounted vertically on one of the side walls and cooled from the back by cooling water (11.5 $^\circ$C) from a commercial cooler (Haake-F3-K, Thermo Fisher Scientific). The sample is back-illuminated with an LED pad (MB-BL305-RGB-24-Z, Metabright) and imaged from the opposite side through an indium-tin-oxide (ITO)-coated heated window with a camera (Point Grey, FL3-U3) through a 20x zoom lens (Z125D-CH12, EHD). The resulting field-of-view is ${\sim}10 \times 7.5$ mm (see movie in Supporting Information \ref{app:movie}). The temperature inside the chamber is measured by several thermistors (TCS651m AmsTECHNOLOGIES and Thorlabs TSP-TH) using a DAQ card and Labview and with the Thorlabs TSP01 Application. Thermistors are located at the vapor inlet, in the vapor close to the sample surface, at the vapor outlet, in the coolant behind the sample, in the heated water on the hot plate, and in the ambient air.

Deionized water (Millipore Synergy UV, 18.2 M$\Omega\cdot$cm) is heated on a hot plate (RCT Basic, IKA labortechnik). Ambient air is blown through the water using an aquarium pump (0886-air-550R-plus, Sera) at a flow rate of of 3.5 l/min, as monitored by a flow meter (AWM5101VN flowmeter, Honeywell). 
The condensation chamber is initially kept dry with a steady flow of dry Nitrogen. At the start of an experiment, the humidified air is guided into the condensation chamber at the bottom of the chamber at a temperature of 42 $^\circ$C, and at a flow rate of 3.5 l/min. To ensure reproducibility, the subcooling of the surface is kept constant at $\sim$30.5 $^\circ$C throughout all experiments. 
 
The recorded images are analyzed using a home-built image analysis routine in MATLAB to evaluate the center locations and radii of all the condensing drops (Supporting Information \ref{app:imageanalysis}). The smallest drop size detectable using this method is $R_{min}\approx 4.3$ $\mu$m.

The interdigitated zigzag electrodes are fabricated using photo-lithography on a glass substrate. The electrodes are subsequently coated with a 2 µm thick dielectric layer of Parylene C (PDS2010, SCS Labcoter) using chemical vapour deposition (CVD), and an ultra thin top hydrophobic polymer coating (CytopTM, Asahi Glass Co., Ltd.) using a dip-coating procedure.  
For the experiments and simulations reported herein, we use interdigitated electrodes with zigzag-shaped edges (Figures \ref{fig:experimentalsetup}b-\ref{fig:experimentalsetup}c). As in ref. \cite{Dey2018}, the minimum and maximum width of the gap between adjacent electrodes are kept fixed at $w_{g,min}=50\mu$m and $w_{g,max}=250\mu$m, and three different lengths $\ell$ of 500, 1000 and 3000 µm are tested. 
For ac-EW, an amplified electrical signal of rms amplitude between $U_{RMS} = 100-150$ V and a fixed frequency of $f = 1$ kHz is used using a function generator (Agilent 33220A) and voltage amplifier (Trek PZD700A). 
Young's contact angle at zero voltage is $\theta_Y\sim110^\circ$ and Lippmann's angle under EW ($U_{RMS} = 150$ V) is $\theta(U_{RMS})\sim90^\circ$.\cite{Baratian2018} The contact angle hysteresis under ac-EW ($U_{RMS} = 100-150$ V) is measured to be $\Delta \cos \theta = 0.06 \pm 0.01$.

\subsection{Numerical Aspects}
To explain our experimental observations, we developed a numerical model that allows us to calculate the electrostatic energy of a drop as a function of its size and the $(x,y)$ position of its center of mass within the unit cell of the electrode pattern (see zoomed view in Figure \ref{fig:experimentalsetup}b). To calculate this energy landscape $(E_{el}(x,y; R))$, we solve the Poisson equation for a three-dimensional computational domain consisting of the electrodes, the dielectric layer, a water drop, and the surrounding air. Since $\theta(150 \ \text{V})\sim90^\circ$, we represent the drop by a simple hemisphere with radius $R$ and with a fixed electrical conductivity ($10^{-5}$ S/m) that guarantees (for all practical purposes) complete screening of the electric field from the inside of the drop. Note that this hemispherical approximation neglects slight EW-induced distortions of the drop shape (see below). Yet, earlier simulations showed that this merely leads to a minor underestimation of the electrostatic trapping strength for rather weakly deformed drops as in the present experiments.\cite{Cavalli2015}

The calculation of $E_{el}(x,y;R)$ starts with the calculation of the distribution of the electrostatic potential $\phi(x,y,z)$ within a three-dimensional domain that encloses a single drop of radius $R$ at a fixed position as well as the adjacent electrodes. $\phi$ and the free charge density $\rho_e$ are related according to the Poisson equation as
\begin{equation}
\label{eq:poisson}
    \nabla^2 \phi = -\frac{\rho_e}{\epsilon_0 \epsilon}.
\end{equation}
Here $\epsilon_0$ is the permittivity of free space, and $\epsilon$ is the relative permittivity of the computational domain.
$\rho_e$ can be related to the current density $\vec{J}$ using the charge conservation equation as
\begin{equation}
\label{eq:chargeconservation}
    \frac{\partial \rho_e}{\partial t} = - \nabla \cdot \vec{J} = \nabla \cdot \sigma \nabla \phi,
\end{equation}
where $\sigma$ is the electrical conductivity of the computational domain. 
Taking the time derivative of Equation \ref{eq:poisson}, and subsequently substituting Equation \ref{eq:chargeconservation} in it, we get a second order partial differential equation in $\phi$:
\begin{equation}
\label{eq:diffequation}    
    \nabla^2 \dot{\phi} = -\nabla \cdot \left ( \frac{\sigma}{\epsilon_0 \epsilon}\nabla \phi \right ).
\end{equation}
Considering a sinusoidal electrical potential $\phi=\phi_0 \Re \left [ e^{i \omega t} \right ]$, and subsequently, considering its time derivative $\dot{\phi} = \phi_0 \Re \left [ i \omega e^{i \omega t} \right ]$, Equation \ref{eq:diffequation} can be rewritten as
\begin{equation}
    \label{eq:poissonsolution}
    \nabla \cdot \left [ \left ( \epsilon_0 \epsilon - i \frac{\sigma}{\omega} \right ) \nabla \phi \right ] = 0.
\end{equation}

Equation \ref{eq:poissonsolution} is solved numerically in COMSOL Multiphysics (version 5.4) using the finite element method for a fixed voltage (amplitude) of $150$ V and frequency of $1$ kHz. The discretization or element order of modeling domains is varied between quadratic and the fifth-order in order to achieve the desired accuracy. Since the drop size in our experiments varies from a fraction of the width of a unit cell at early stages to drops covering several adjacent electrodes during later stages, the computational domain is chosen to be sufficiently large to cover the entire drop as well as the immediately adjacent electrodes. (In practice, we chose several domain sizes for different ranges of drop sizes in order to reduce computational efforts.) The geometries of electrodes and dielectric films are chosen according to the experiments. Dirichlet boundary conditions (fixed electrostatic potential) are imposed on the electrode surfaces; Von Neumann conditions (zero electric field in normal direction) are applied on all other boundaries. Supplementary Information \ref{app:numerical_geometry} shows a typical view of a computational domain along with the resulting potential distribution for a specific drop configuration. As mentioned above, these calculations were repeated for 200 values of the drop size R between 0 and 900 $\mu$m, and for each drop size at $30 \times 30$ (large R) or $30 \times 60$ (small R) equally spaced locations within the unit cell. (For symmetry reasons, it is sufficient to vary the drop positions only within half of a unit cell; see grey shaded area in Supplementary Information \ref{app:numerical_geometry}.)
 
After numerical evaluation of $\phi(x,y,z)$ for all allowed drop sizes and  $(x,y)$-location within the unit cell, the total electrostatic energy of the entire system is calculated as
\begin{equation}
    \label{eq:Eel}
    E_{el}(x,y;R) = - \int_v \frac{1}{2} \vec{E} \cdot \epsilon_0 \epsilon \vec{E} dv = -\frac{1}{2} \int_v \epsilon_0 \epsilon \left ( \left | \frac{\partial \phi}{\partial x} \right |^2 + \left | \frac{\partial \phi}{\partial y} \right |^2 + \left | \frac{\partial \phi}{\partial z} \right |^2 \right ) dv,
\end{equation}
where $\vec{E} = - \nabla \phi$ is the electric field, and the integration represents the volume integral over the entire computational domain. 

In the representation of the electrostatic energy landscapes later on (Figure \ref{fig:model}), we make use of symmetries and periodicities to extend the energy landscapes beyond a single unit cell for a more intuitive representation.
Finally, note that Equation \ref{eq:poissonsolution} contains both dielectric and purely conductive contributions.
However, for the conductivity of pure water and for the applied (low) frequency, the ionic current dominates the displacement current towards screening the electric field (also see \cite{Baratian2018}). 

\section{Results}

\subsection{Evolution of breath figures}

 \begin{figure}[t]
\centering
  \includegraphics[width=1\linewidth]{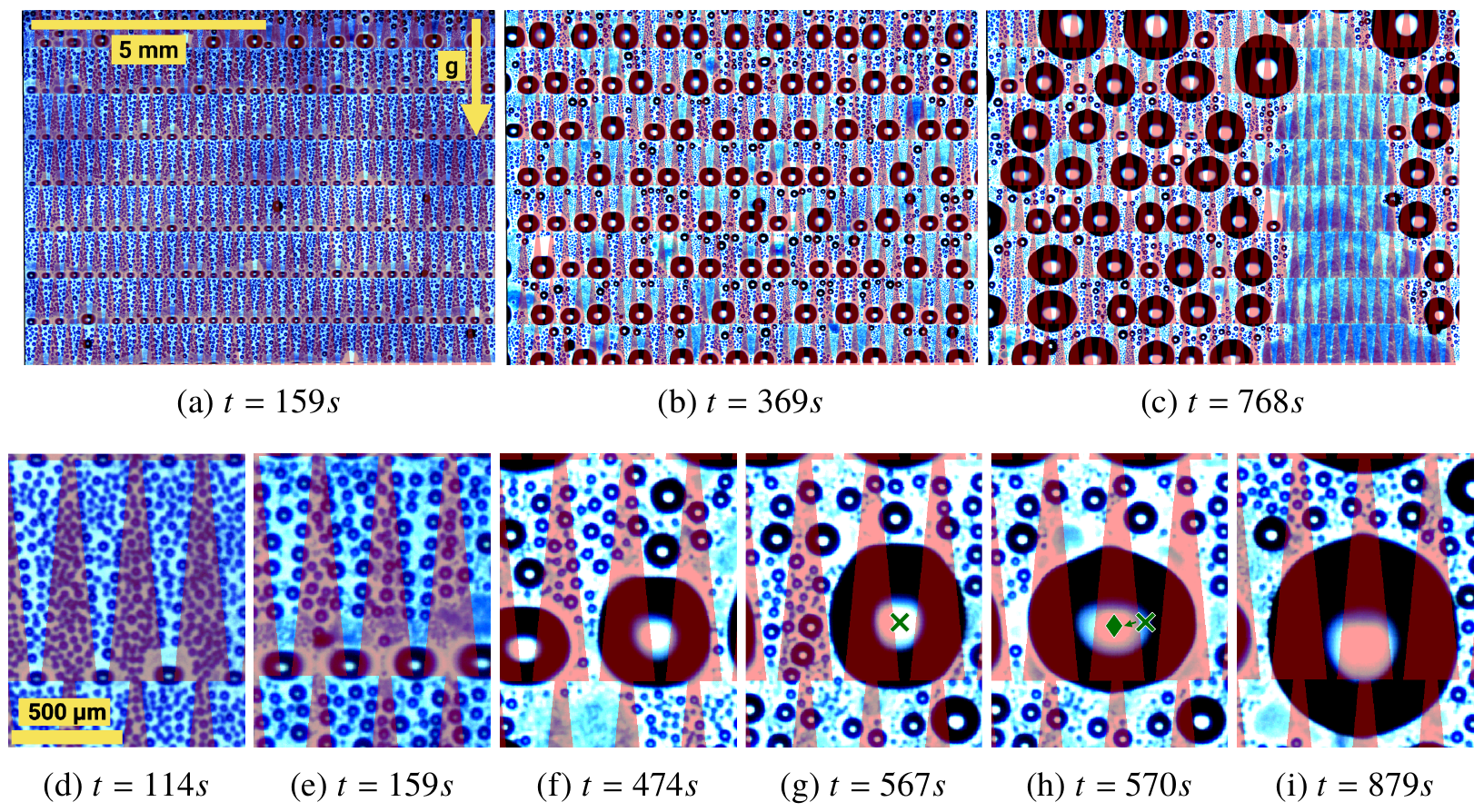}
\caption{Top view of condensed droplets on a vertically mounted substrate ($l=1000\mu$m). (a-c) Full field of view for $t=159, 369, 768$s illustrating alignment and growth of condensing drops. (d-i) Zoomed view of $\sim$3 unit cells for times as indicated. Note the vertical and horizontal shift of the center of the drops with increasing size. (Transparent ITO electrodes are superimposed in red.)}
\label{fig:breathfigures}
\end{figure}

 As apparent at first glance, the condensate drops form a pattern (breath figure) with well-defined periodicities along both the lateral $(x-)$ and vertical $(y-)$ direction upon condensation onto surfaces with zigzag interdigitated electrodes (Figures \ref{fig:breathfigures}a-\ref{fig:breathfigures}c). 
 This is in sharp contrast to breath figures with straight interdigitated electrodes under ac-EW, where no periodicity along the $y-$direction was found.\cite{Baratian2018} While these observations have qualitatively been reported before,\cite{Dey2018} a closer look at the representative Figures \ref{fig:breathfigures}d-\ref{fig:breathfigures}i reveals a number of additional details: 
 Initially, the small condensate drops are essentially randomly distributed; however, as the drops grow and begin to coalesce, they align parallel to the electrode edges, with a slight preferential displacement towards the gap centers (Figures \ref{fig:breathfigures}d-\ref{fig:breathfigures}e).
 Simultaneously, the drops closer to a gap minimum (i.e. at $(x=w/2; y=0)$ in Figure \ref{fig:experimentalsetup}b) are pulled down towards that minimum and typically grow on their way by coalescence with other drops (Figures \ref{fig:breathfigures}d-\ref{fig:breathfigures}e). 
 As we will see below, drops at these `gap minima' are trapped in electrostatic energy minima; as these continue to grow, their lower edge remains close to $y=0$, whereas their center gradually moves upwards (Figures \ref{fig:breathfigures}d-\ref{fig:breathfigures}g). 
 These trapped, growing condensate drops dominate the visual appearance of the breath figures on a macroscopic scale (Figures \ref{fig:breathfigures}a-\ref{fig:breathfigures}b).
 Interestingly, upon reaching some critical size, the center of mass of these trapped drops suddenly translates horizontally from being centered on the gap between two adjacent electrodes to being centered on an electrode (see transition cross marker to diamond marker in Figures \ref{fig:breathfigures}g-\ref{fig:breathfigures}h).  
Upon growing further, the center of the trapped drops shifts slightly downwards (Figure \ref{fig:breathfigures}i). 
Eventually, at much larger radii, drops shed under the influence of gravity when the individual drop weight exceeds the electrical trapping force and contact angle hysteresis (Figure \ref{fig:breathfigures}c).

\begin{figure}[ht]
\centering
  \includegraphics[width=1\linewidth]{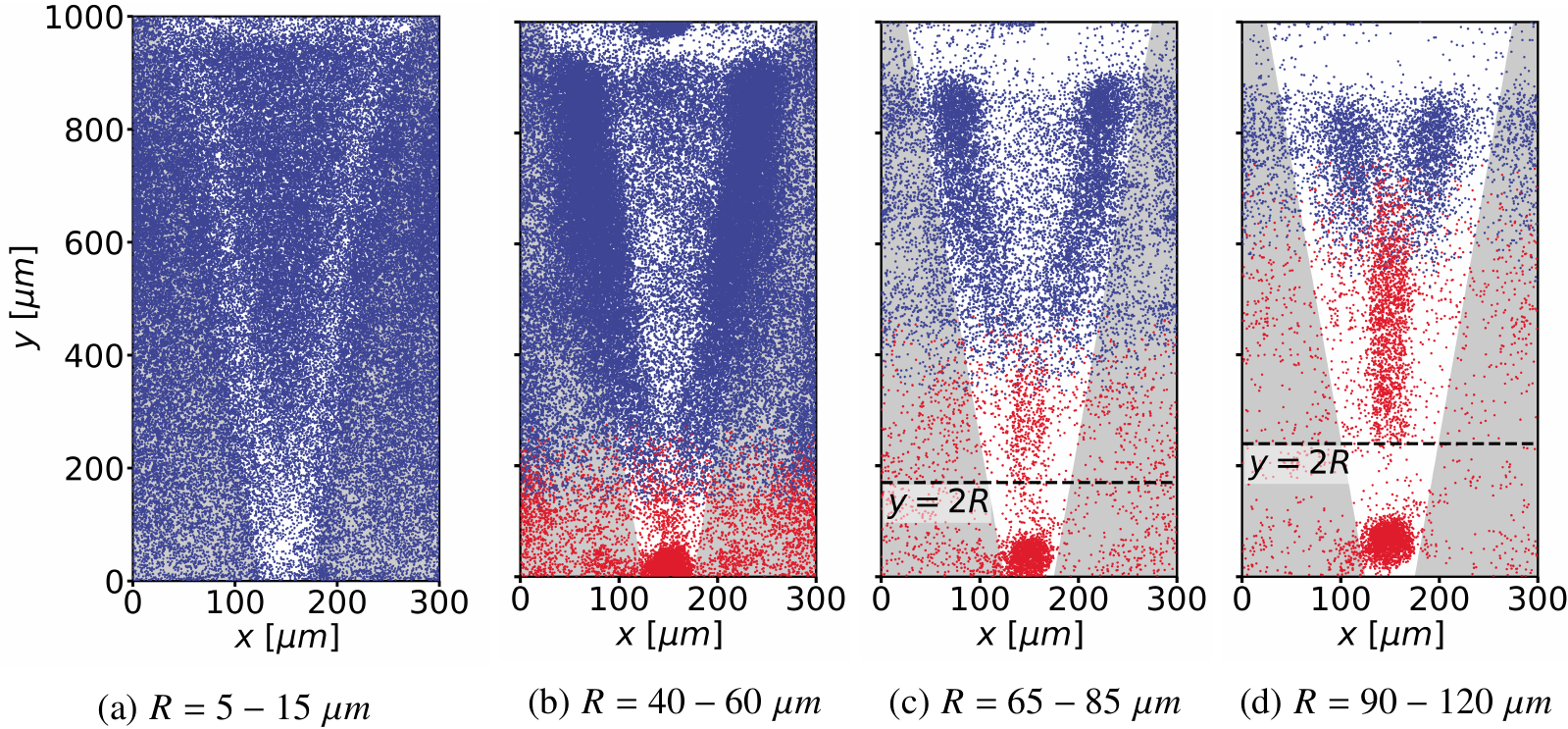}
\caption{Center locations of all drops projected into single unit cell and binned to  size ranges as indicated ($l=1000\mu$m). Drops with $R>0.5w_g(y)$ are shown in red. $w_g(y)$ is the $y$-dependent gap width ranging from $w_g(0)=w_{g,min}$ to $w_g(\ell)=w_{g,max}$.}
\label{fig:spatialdistributions}
\end{figure}

For a statistical analysis of the distribution of condensate drops , we project the drop centroid locations of all drops within the field of view onto a single unit cell of the electrode pattern (Figure \ref{fig:experimentalsetup}b) using a mapping procedure that takes into account the optical distortion of the optical imaging system (see Supporting Information \ref{app:imageanalysis}). 
Figure \ref{fig:spatialdistributions} shows the resulting spatial distribution of the drops within the unit cell for $\ell = 1000$ $\mu$m binned into ranges of $R=5-15, 40-60, 65-85$, and $90-120$ µm, where each data point represents the location of a drop center at a particular moment in time.

While the distribution of the smallest drop sizes (5-15 µm; Figure \ref{fig:spatialdistributions}a) is almost random, somewhat larger drops (40-60 µm) preferentially align along the inclined edges of the electrodes (Figure \ref{fig:spatialdistributions}b). 
As the drops coalesce and grow further, they gradually move from the electrode edges towards the gap center (Figures \ref{fig:spatialdistributions}b-\ref{fig:spatialdistributions}d). 
Drops with a diameter that exceeds the local width $w_g(y)$ of the gap, i.e. drop with a critical size $R>0.5w_g(y)$ (red data points) are preferentially found in the center of the gap rather than along the electrode edges (Figure \ref{fig:spatialdistributions}c-\ref{fig:spatialdistributions}d), giving rise to a peculiar bi-modal distribution of the drops (Figure \ref{fig:spatialdistributions}c). This bi-modal spatial distribution of drops is unique to the converging electrode geometry. 
In contrast, for straight electrode edges with a constant gap width, the drop distribution is always uni-modal (i.e. the drops of equal size align either on both sides of the gap center or along the gap center).\cite{Baratian2018}
The larger the drop size under consideration, the larger the fraction of drops with a width exceeding the local gap width, i.e. with $R>0.5w_g(y)$. 
Hence, the largest drops are again largely centered on the gap (red dots in Figure \ref{fig:spatialdistributions}d), concomitant with a depletion of drops from the electrodes including their edges.
The evolution of the spatial distribution of the condensate drops (Figures \ref{fig:spatialdistributions}b-\ref{fig:spatialdistributions}d) with increasing drop size is thus reminiscent of a `zipper-like' effect.
The cluster of data points in the vicinity of the gap minimum always represent electrically trapped droplets (Figures \ref{fig:spatialdistributions}b-\ref{fig:spatialdistributions}d).
Furthermore, Figures \ref{fig:spatialdistributions}c-\ref{fig:spatialdistributions}d clearly show that the strong electrical force sweeps the drops within a distance of characteristic length scale ${\sim}2R$ above the gap minimum (note the relative lack of droplets over this region) creating the bigger trapped droplet which continues to grow upward.

\begin{figure}[ht]
\centering
 \includegraphics[width=1\linewidth]{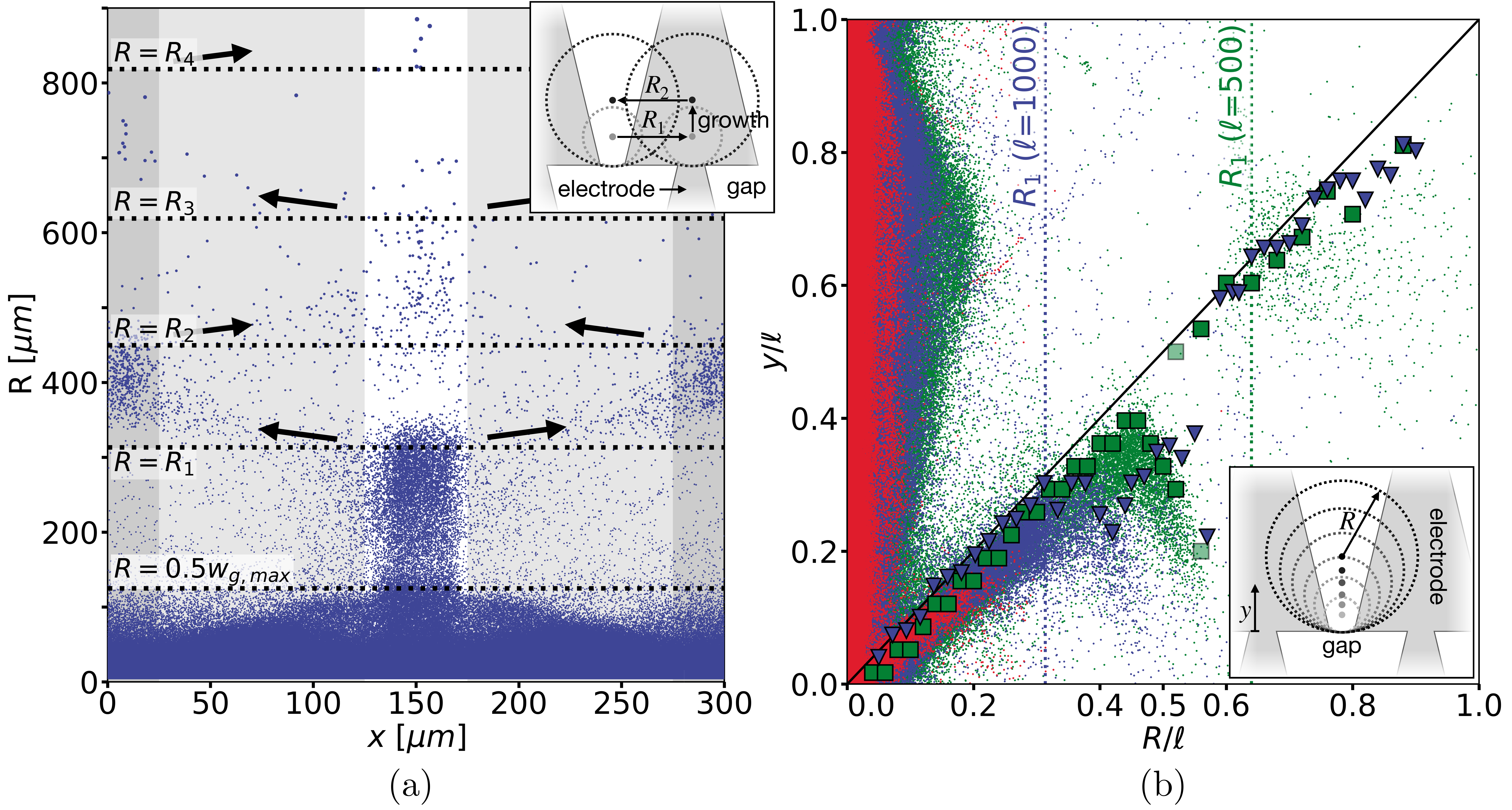}
\caption{Horizontal (a) and vertical (b) distribution of experimental drop centers (small dots) within a unit cell for variable drop radius $R$. (a) Horizontal dotted lines indicate critical radii for first central alignment (0.5$w_{g,max}$) and subsequent transitions ($R_1...R_4$) from preferred alignment on gap center ($x=150\mu$m) and electrode center ($x=0$ or $300\mu$m), as extracted from Figure \ref{fig:modeltransitions} (data for $l=1000\mu$m). Bright and light shaded regions indicate the minimum and maximum gap widths $w_{g,min}$ and $w_{g,max}$. Inset: illustration of horizontal transitions of drop center upon growth. (b) Normalized vertical position of drop center for all electrode sizes (green: $l=500\mu$m; blue: $l=1000\mu$m; red: $l=3000\mu$m) vs. normalized drop size. Large symbols: vertical position of electrostatic energy minimum vs. drop size extracted from numerical calculations (see Figure \ref{fig:model}): green squares: $l=500\mu$m; blue triangles: $l=1000\mu$m. Solid line: geometric approximation $y=R$. Inset: illustration of geometric shift of drop center for bottom of drop pinned at minimum gap.}
\label{fig:statisticalalignment}
\end{figure}

Another interesting series of transitions is revealed by plotting the correlation between average drop size $R$ and the lateral position of their center of mass (Figure \ref{fig:statisticalalignment}a). For $R>0.5w_{g,max}$, most drops are preferentially aligned along the gap center ($x=150$ $\mu$m). 
However, this gap-centered alignment of the condensate drops does not persist as the drops grow further. At another critical size $R_1$ (${\sim}320$ $\mu$m for the present $\ell=1000$ $\mu$m electrode), drops on average undergo a transition from being centered on the middle of the gaps to being centered on the middle of the adjacent electrodes, as already described for the specific individual drop in Figures \ref{fig:breathfigures}g-\ref{fig:breathfigures}h. Beyond that, a series of additional transitions back and forth the centers of gaps and electrodes are seen at critical radii $R_2, R_3, R_4$, yet increasingly faint due to decreasing numbers of larger drops. The positions of the dashed horizontal lines emerge from the numerical model (see below). The same series of transitions are also observed for the other electrode geometries with $\ell=500$ $\mu$m and $\ell=3000$ $\mu$m (see Supporting Information \ref{app:horizontaltransitions_extra}).

In order to further visualize the spatial evolution of the trapped drops, Figure \ref{fig:statisticalalignment}b shows the projected vertical ($y$-) locations of all drops normalized by the electrode length $(y / \ell)$ versus $R/\ell$ for all three electrode designs ($\ell=500, 1000, 3000$ $\mu$m).  
The tail developing from the gap minimum $(y/\ell=0)$ represents the vertical locations of the trapped drops.
Initially, the vertical locations of these trapped drops satisfy $y\approx R$ (solid black line in Figure \ref{fig:statisticalalignment}b), as previously observed from Figures \ref{fig:breathfigures} and \ref{fig:spatialdistributions}.
Although for the $\ell=3000$ $\mu$m electrode (red data points) the trapped drops stay aligned at $y\approx R$ till gravity-driven shedding ($R_{shed}/\ell \approx 0.3 $), for the $\ell=1000$ $\mu$m (blue) and $\ell=500$ $\mu$m (green) electrodes the drops subsequently deviate from the $y=R$-trend as these continue to grow by coalescence (Figure \ref{fig:statisticalalignment}b).
For the $\ell=1000$ $\mu$m electrode, this deviation begins immediately after the first lateral transition ($R_1/\ell \approx 0.32$), while for the $\ell=500$ $\mu$m electrode this occurs well before the first lateral transition at $R/\ell \approx 0.5$ (Figure \ref{fig:statisticalalignment}b). 
Interestingly, once the trapped drops grow bigger, they realign again following $y\approx R$ for both the $\ell=1000$ $\mu$m and $\ell=500$ $\mu$m electrodes above $R/\ell \approx 0.6$ (Figure \ref{fig:statisticalalignment}b).
Note that for both $\ell=1000$ $\mu$m and $\ell=500$ $\mu$m electrodes, the deviation of the trapped drops from the $y = R$ trend occurs well below the critical shedding radius ($R_{shed}\approx1$ mm); hence, gravity is not the cause of the deviation.

\subsection{Electrostatic energy landscape controls the evolution of breath figures}
\begin{figure}[!htbp]
	\centering
	\includegraphics[width=1\linewidth]{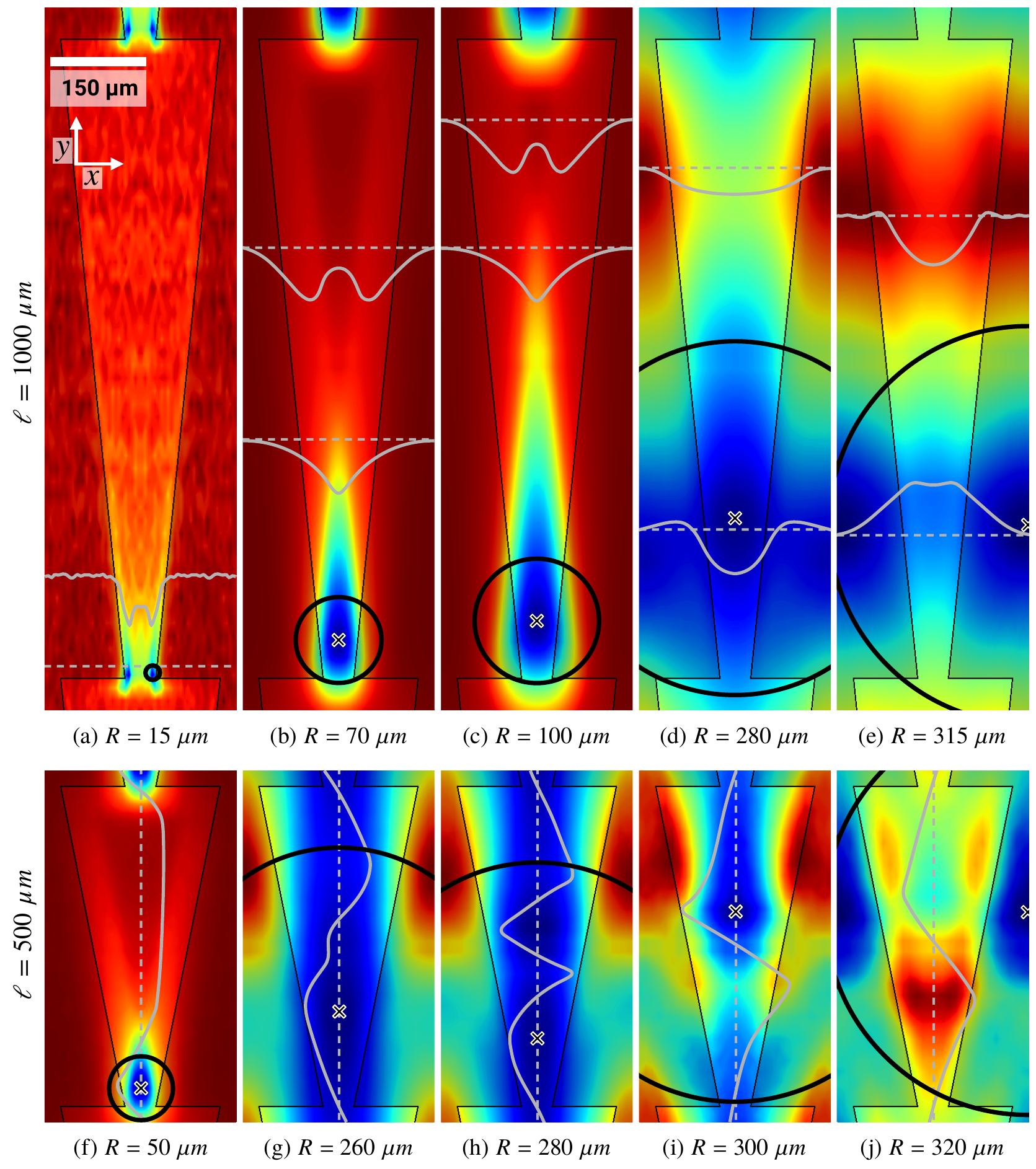}
	\caption{False color representation of 2D energy landscape (blue=low; red=high energy) vs. $(x,y)$ position of drop center for various drop sizes as indicated. Top row (a)-(e): $l=1000\mu$m; bottom row (f)-(j): $l=500\mu$m. Grey lines: cross sections through electrostatic energy landscapes along dashed lines; crosses and black circles: centers and edges of drops at minimum energy configuration.(Color scales and cross sections are individually rescaled for optimum visual contrast; Data in a) are somewhat compromised by numerical noise.)}
	\label{fig:model}
\end{figure}

The details of the drop distributions described above can be understood by considering the 2D electrostatic energy landscape ($E_{el}(x,y)$) emerging from our numerical calculations. Figure \ref{fig:model} illustrates the evolution of these energy landscapes for electrodes with $\ell=1000$ $\mu$m Figures \ref{fig:model}a-\ref{fig:model}e) and $\ell=500$ $\mu$m (Figures \ref{fig:model}f-\ref{fig:model}j) for a series of drop sizes as indicated in the figure. As noted above, the energy landscape is by construction mirror symmetric along the center of the gap. For the smallest drops, the energy landscape is rather flat with shallow valleys along the electrode edges that become deeper upon approaching the gap minimum at $y=0$ (Figure \ref{fig:model}a). For larger drops, (Figure \ref{fig:model}b, \ref{fig:model}c), the two separate minima along the electrode edges first merge into one minimum centered at $x=w/2, y\approx R$, leading to a coexistence of a single minimum close to the gap center in the lower parts of the unit cell and two valleys in the upper parts. Such variation in the $E_{el}(x,y)$ landscape is  consistent with the drop distribution shown in Figure \ref{fig:spatialdistributions}b-\ref{fig:spatialdistributions}d and its 'zipper-like' evolution with increasing drop size. The $y-$coordinate of the central minimum gradually shifts towards larger $y$ for sizes comparable to the trapped drops, consistent with the solid $y=R$ line in Figure \ref{fig:statisticalalignment}b.

\begin{figure}[!htbp]
\centering
  \includegraphics[width=1\linewidth]{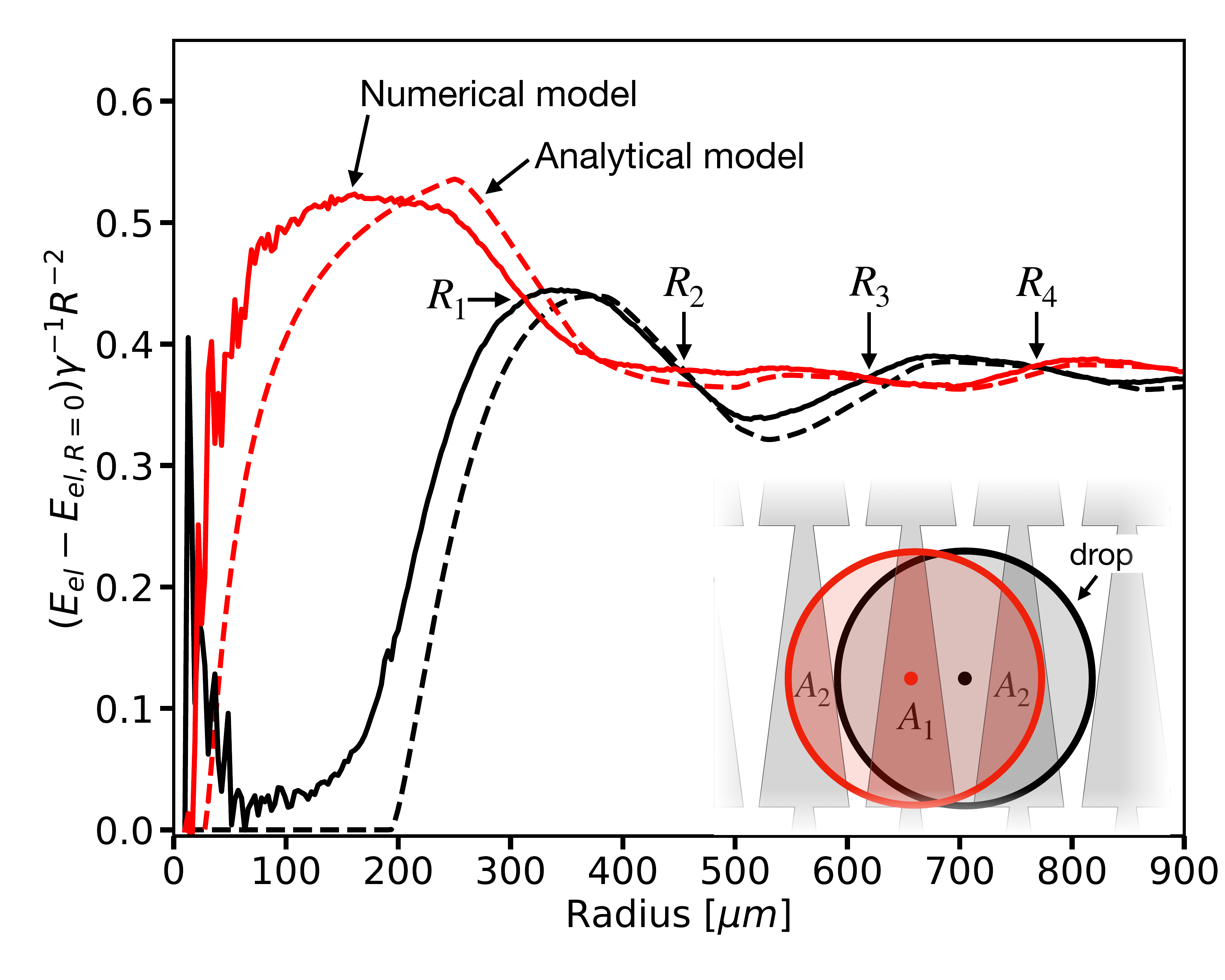}
\caption{Normalized excess electrostatic energy vs. drop size for drops centered on the gap (black) vs. centered on the electrode (red). Solid: full numerical model; dashed: analytical model based on drop-electrode overlap areas $A_1, A_2$. $R_1, ... R_4$: critical radii of transitions from preferential gap center to electrode center-alignment. Inset: illustration of competing drop positions.}
\label{fig:modeltransitions}
\end{figure}

 As the drop size approaches 300 $\mu$m for $\ell=1000$ $\mu$m, the electrostatic energy minimum eventually moves from the gap center to the electrode center (Figures \ref{fig:model}d-\ref{fig:model}e). 
We can predict the drop radius for this and subsequent lateral transitions by calculating the $E_{el}$ for a drop located either at the gap center or at electrode center, for a range of drop radii (Figure \ref{fig:modeltransitions}). For small drop sizes ($R<320 \ \mu$m), the total electrostatic energy is smaller when the drop is located at the gap center (black solid line in Figure \ref{fig:modeltransitions}) than when located at the electrode center (red solid line). As the drop size increases further, the location of the lowest electrostatic energy moves alternately between the electrode and the gap centers (compare the relative variations between the black and red solid lines in Figure \ref{fig:modeltransitions}). The characteristic radii at which these transitions occur, $R_1$, $R_2$, ..., $R_n$ in Figure \ref{fig:modeltransitions} are shown as horizontal dashed lines in Figure \ref{fig:statisticalalignment}a, and provide a good description of the transitions observed experimentally. Note that these horizontal transitions of the center of mass as a function of the drop size are well-known for surfaces with parallel stripes of alternating wettability originating from chemical patterning.\cite{Brandon1997}

Tracing the position of the global energy minimum along the $y-$direction reveals that the drop center indeed moves upward with increasing drop size following slightly below the line $y=R$, as shown in Figure \ref{fig:statisticalalignment}b. The numerical results (squares and triangles in Figure \ref{fig:statisticalalignment}b) reproduce the experimental observations with great accuracy. In some cases correlations with lateral transitions of the drop position can be observed. 

For the shorter unit cell ($\ell=500 \ \mu$m), the evolution of the energy landscape is qualitatively similar. Nevertheless, the two situations cannot be mapped directly onto each other. For instance, unlike the long electrodes, we find for $\ell=500 \ \mu$m that the energy minimum in the gap center splits up into two distinct local minima as the drop diameter becomes comparable to $\ell$ between $R\sim250 \dotso \sim300 \ \mu$m (Figures \ref{fig:model}g-\ref{fig:model}i). This leads to a distinct transition of the drop position along the $y-$direction for $R=280 \rightarrow 300 \ \mu$m (Figures \ref{fig:model}h-\ref{fig:model}i), while the drop remains laterally centered on the gap. This transition is indeed observed in the experiments with short ($\ell=500 \ \mu$m) electrodes (Supplementary Information \ref{app:horizontaltransitions_extra}a) but not for $\ell=1000 \ \mu$m, see Figure \ref{fig:statisticalalignment}a. Nevertheless, the slight downward shift of the center-off-mass position for $R/\ell \sim 0.4...0.5$ in Figure \ref{fig:statisticalalignment}b is also correctly reproduced for electrodes of both short and intermediate length. 

As an alternative to the full numerical calculations, we can also evaluate the energy landscapes by approximating the electrostatic energy using the simple geometric approximation proposed by 't Mannetje et al.\cite{Mannetje2013,TMannetje2014}
This analytical calculation involves approximating the condensate drop-dielectric system as an electrical circuit consisting of two parallel plate capacitors in series formed by the overlap between the conducting drop and the electrodes. The overall capacitance of the system is approximated as $C(x,y) = \epsilon_0 \epsilon_d / d \cdot A_{cap}$, where $A_{cap}= A_1 A_2 / (A_1 + A_2)$, and $A_1$ and $A_2$ are the spatially varying overlap areas between the drop footprint and the two electrodes (see inset in Figure \ref{fig:modeltransitions}). The associated electrostatic energy of the system on application of an electrical voltage $(U)$ can be written as $E_{el,cap}=-C(x,y) U^2 / 2$.\cite{Mannetje2013} The dashed lines in Figure \ref{fig:modeltransitions} show the electrostatic energy minimum in this approximation for drop centered on the gap (black) and on the electrode (red). Like in the case of the full numerical model, for small drops (i.e. $R<\sim320 \ \mu$m) it is more favorable to be centered on the gap, whereas for increasing $R$ there is a succession of transitions between preferred alignment on the electrode center and the gap center. While the energies deviate substantially for the smallest drop sizes (for which the overlap area with one of the electrodes and hence the total energy can vanish), the agreement improves for increasing drop size and the predictions for the various subsequent transitions of the drop positions ($R_2, R_3, R_4$) becomes remarkably good. 

While some of the aspects described above are very specific to the present electrode configuration, the overall excellent agreement demonstrates the ability of the numerical model to reproduce the experiments, including even subtle aspects such as the transitions between various competing local minima of the overall energy landscape are correctly captured. For not too small droplets, the simple analytical model of geometric overlap also provides reasonable predictions between various competing drop configurations. 

\section{Discussion and Perspectives}
The results presented here clearly demonstrate the flexibility of electric fields in controlling condensation patterns on solid surfaces with submerged co-planar electrodes. While individual drops are obviously subject to their specific local environment, averaging over large ensembles shows that condensed drops decorate the local minima of electrostatic energy landscapes of remarkable complexity. Consequently, the drops undergo gradual translations as well as discrete transitions as local shift or become unstable with increasing drop size. Apparently, the random character of coalescence events with neighboring drops in combination with enhanced mobility of drops caused by the reduced contact angle hysteresis in EW with AC voltage\cite{Li2008} provide sufficient energy for the drops to explore the entire energy landscape despite the fact that energetic barriers between adjacent minima are obviously substantially larger than thermal energies. While not exploited here explicitly, compared to passive chemical or topographic patterning, EW-functionalization offers the advantage of switchability in addition to the enhanced drop mobility thanks to the reduced effective hysteresis. 

While drop positions are well-defined and controllable beyond a certain critical size, the random distribution of small drops in Figure \ref{fig:spatialdistributions}a confirms the earlier observation that the nucleation sites for drop condensation seem to be unaffected: no correlation can be observed between the position of the smallest drops and the location of the electrodes. This arises from the fact that the forming liquid nuclei only experience a dielectrophoretic polarization force. Upon nucleation, this electrostatic force competes with surface forces caused by random heterogeneities on the surface. While the electric force scales with the (very small) volume of the critical nucleus, i.e. $\propto R^3$, the latter scale with surface area, i.e. $\propto R^2$ and therefore dominate. A control of nucleation rates and locations is therefore possible only if local electric fields and field gradients can be substantially increased, e.g. by generating miniaturized electrode patterns on the nanoscale. 

From an applied perspective, the key question in both fog harvesting and enhanced heat transfer is how the removal of drops from the surface can be optimize to condense as much liquid as possible. Obviously, this requires a somewhat broader perspective of the entire system than only the control of drop distribution patterns. While the results presented above clearly show that suitable electrode patterns allow to control drop positions and to promote faster growth by inducing lateral and vertical translations and coalescence, the same strong electrostatic forces also generate deep energetic traps that can hold back even large drops as shown in Figure \ref{fig:breathfigures} and thereby hamper efficient drop removal. To circumvent this problem, the electrically induced wettability patterns should be applied with some form of time-dependent actuation. The easiest approach is to periodically activate the electrodes to induce drop motion and growth, and to subsequently deactivate them such that drops exceeding the relevant critical size can spontaneously shed off the surface under the influence of gravity. While some success of this strategy has been demonstrated,\cite{Dey2018,Wikramanayake2020} the overall performance was not impressive. In part, this is probably caused by the fact that the pinning forces increase as the EW-induced reduction of the contact angle hysteresis ceases upon switching off the AC voltage and hence the critical shedding radius increases and the shedding frequency decreases.\cite{Mannetje2011a} Alternatively, one could make use of active transport strategies borrowed from EW-based lab-on-a-chip systems, where drops are transported towards activated electrodes.\cite{MugeleBook} Given the nature of drop condensation, it is obviously not desirable to bring the condensing drops in direct contact with electrodes on top of the functionalized surface. Therefore, structured electrodes, possibly in two layers, should be embedded into the substrate and actuated in such a manner that they lead to a conveyor belt-like directed motion. Such strategies are rather straightforward to implement for surfaces that are flat or covered by some `moderate' degree of topographic pattern. For intrinsically three-dimensional structures such as meshes that are frequently used for fog harvesting the implementation of any form EW-enhanced condensation and drop removal is much more difficult to realize - notwithstanding initial demonstrations with crossing fibers of switchable wettability.\cite{Eral2011} \\

While the effect of EW on the drop distribution patterns is rather striking, the reported consequences for the total condensation rate and the resulting heat transfer are far less impressive.\cite{Baratian2018,Dey2018,Wikramanayake2019} Applying standard models of dropwise heat transfer,\cite{Zhao2018a} Wikramanayake et al. pointed out that the majority of previous EW experiments were not very enlightening because they were carried out using water vapor in moist air.\cite{Wikramanayake2019} Under such conditions, it is well-known in the heat transfer community that the overall heat transfer resistance is dominated by the ambient air, which acts as a non-condensable background gas and introduces a diffusive boundary layer at the solid-vapor interface.\cite{Minkowycz1966} The expected beneficial effects of EW-enhanced drop removal on heat transfer are thus overshadowed by mass transport limitations across that boundary layer and thus the actual potential of the EW-induced enhancement does not become evident. To demonstrate and exploit the benefits of EW, condensation setups should thus be designed in such a manner that drop removal is indeed the dominating factor for the overall heat transfer coefficient. This implies in particular preferential operation in pure vapor. \\
  
A final essential issue for any practical application of the effects described above is the stability of the EW-functionalized surfaces over extended periods of time. Like many other applications, both fog collection and heat transfer require long continuous operation times of the devices, ideally of the order or years. While proof-of-principle experiments in laboratories on short time scales are often relatively easy to achieve, maintaining cleanliness and hydrophobicity of coatings in the presence of complex and reactive fluids such as condensing water vapor is extremely demanding from a materials perspective. Recent experiments demonstrated that fluoropolymer surfaces commonly used in EW spontaneous charge up upon contact with water for several hours.\cite{Banpurkar2017} In the presence of electric fields, this effect is even more pronounced and can even be exploited to generate well-controlled charge densities and charge patterns.\cite{Wu2020,Wu2020a} Therefore, the development of reliable hydrophobic fluoropolymer coatings that remain stable throughout the life time of various types of devices has been a long standing challenge in applied EW research. One recommendations from these investigations has been to avoid water as an operating fluid whenever possible.\cite{Raj2009} While this is an interesting option for heat transfer devices, it is obviously not possible for fog harvesting applications. In such cases, novel materials with improved stability\cite{Paxson2014} will be required to achieve the necessary stability of operation. Nevertheless, it is also worth pointing out that extended EW-enhanced drop condensation tests over a period of 40 hours displayed - after some initial degradation within the first 1-2 hours - rather stable operation even for conventional fluoropolymer surfaces (Supporting Information \ref{app:surfacedegradation}). 

\section{Conclusions}
Co-planar electrodes embedded into electrowetting-functionalized surfaces allow to control the distribution of drops of condensing water vapor. For interdigitated electrodes with zigzag-shaped edges drops undergo a series of transitions between different preferred locations as they grow in size upon further condensation and coalescence. Comparison to numerical calculations shows excellent agreement with the experiments and demonstrates that the drops decorate on average the minima of the drop size-dependent electrostatic energy landscape, including subtle transitions between preferred locations. This agreement demonstrates that the existing numerical approach provides a solid basis for future more sophisticated models that will include time-dependent electrical actuation schemes. Such models can eventually be used by engineers for numerical optimization of electrode designs and operation modes of future EW-enhanced drop condensation systems. A critical assessment of bottlenecks for such applications indicates that mass transfer limitations and - in particular - the development of combinations of long term stable condenser surface materials and fluids will be essential for the technological success of the approach. 

\begin{acknowledgments}
We thank D. Wijnperlé for sample preparation in the cleanroom and D. Baratian for his contributions in earlier phases of the experiment. We acknowledge financial support by the Dutch Organization for Scientific Research (NWO) within the VICI program (Grant No. 11380).
\end{acknowledgments}

\appendix

\section{Movie of condensation process}
\label{app:movie}

This video: \url{https://drive.google.com/file/d/1RWa03Sf3_Wf9_bE84jdU-lzbduahjOy4/view?usp=sharing}

\section{Image analysis and data processing}
\label{app:imageanalysis}

\subsection{Image analysis algorithm}
\label{app:imageanalysis:imagenanalysis}
The high contrast color images recorded of the condensation process are loaded onto a computer for analysis using a home-made program in Mathworks MATLAB. The goal of the image analysis is to extract the exact location and size of every drop in the image and from there estimate the total volume on the surface and condensation rate.

\begin{figure}[!htbp]
\centering
  \includegraphics[width=1\linewidth]{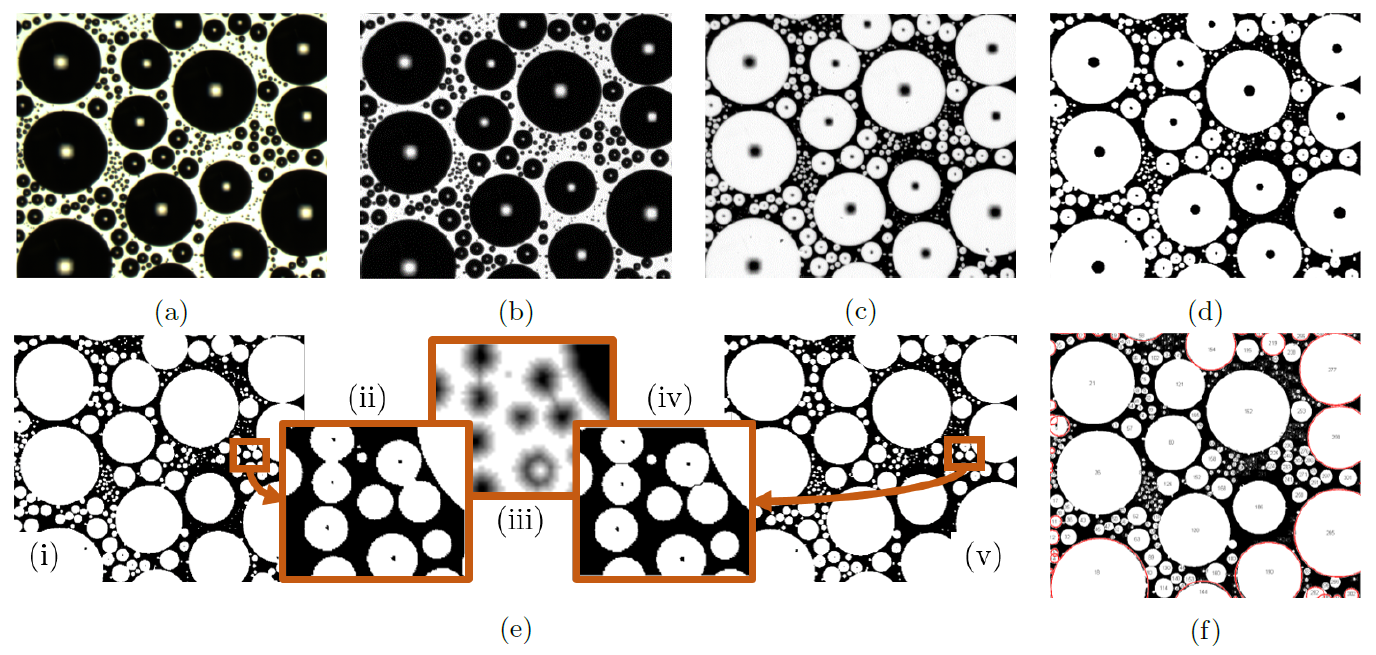}
\caption{The basic image analysis procedure for determining the location and size of every drop. The original image (a) is converted to grayscale (b). The background image is subtracted and the contrast is increased (c). The image is converted to a binary image (d), holes inside drops are filled and connecting drops are separated (e). Holes in drops are filled, the separate components are detected and the boundary drop detection is applied (f).}
\label{fig:imageanalysis}
\end{figure}

A shortened version image analysis steps is shown in Figure \ref{fig:imageanalysis}. First the original image (Figure \ref{fig:imageanalysis}-(a)) is converted to a gray-scale image (Figure \ref{fig:imageanalysis}-(b)). Then a background image, that is an image with no drops on the surface recorded before a condensation experiment, is subtracted from it and the contrast is increased (Figure \ref{fig:imageanalysis}-(c)). Otsu’s method is used to automatically determine the binary threshold to convert the image to a binary image. The Otsu’s method determines the binary threshold in a manner such that the combined spread (variance) of black and white pixels is minimal. Using the obtained threshold, each image is then converted to a binary image, see Figure \ref{fig:imageanalysis}-(d). In order to remove the holes inside the drops the separate background components (regions of connecting black pixels) are detected. Each connected component with a threshold eccentricity and radius are filled (inset Figure \ref{fig:imageanalysis}-(e)-(i)). Filling based on the eccentricity and radius prevents filling connected drops. Aforementioned, some drops are slightly connected in the binary image (inset Figure \ref{fig:imageanalysis}-(e)-(ii)). Separating these drops starts by taking the distance transform of the image where each foreground (white) pixel is transformed into the closest distance to a background (black) pixel. The resulting image (inset Figure \ref{fig:imageanalysis}-(e)-(iii)) shows a gradient from white to black where the darker the color, the further that pixel is inside a drop. The tiny local minima are removed and the watershed transform is applied to the distance transform and translated to the binary image. The watershed transform turns local minima in the foreground pixels (the shortest connection between two connection drops) into background pixels, thus splitting the connected drops (inset figures \ref{fig:imageanalysis}-(e)-(iv) and \ref{fig:imageanalysis}-(e)-(v)).

The non-circular holes in the drops are filled and the connected components of the drops are determined. Each component now represents one drops. For drops not touching the border of the image (non-boundary drops) is assumed that they are circular. Assuming a spherical projection of the drops on the surface the radius and center of each drop is calculated.

Drops touching the border of the image (boundary drops) undergo an extra analysis step. Each of the components contours are circle fitted from which the radius and center location is determined. The volume inside the frame is determined based on the radius, the part of the drop inside the image and the assumption that the drop is a hemisphere.

The end result of the image analysis is shown in Figure \ref{fig:imageanalysis}-(f) where each drop is numbered uniquely.

Now that the location and center of each drop is known, data can be extracted from the breath figures. This data includes the radius of shedding drops, surface coverage and total volume in the breath figure. The total volume assumes a constant drop contact angle of $115^\circ$ to calculate the volume of the spherical caps inside the frame. For drops touching the edge of the frame, the volume inside the frame is estimated assuming hemispherical drops minus (half) spherical caps.

\subsection{Image lens distortion correction}
\label{app:imageanalysis:distortion}
The recorded images of the breath figure evolution are slightly pincushion distorted. Projecting the data onto a single small unit cell will result in a spread in the statistical analysis. To correct for a lens distortion correction is applied to analyzed data.

Lens distortion can be expressed as:\cite{Wang2008}
\begin{align}
\label{eq:1a}
    u & = u_d + \delta_{ud}(u_d,v_d) \\
    \label{eq:1b}
    v & = u_v + \delta_{vd}(u_d,v_d) 
\end{align}
where $u$ and $v$ are the unobservable distortion-free image coordinates; $u_d$ and $v_d$ are the corresponding image coordinates with distortion; $\delta_u(u,v)$ and $\delta_v(u,v)$ are distortion in $u$ and $v$ direction, respectively.\\
Distortion can be classified in three categories: radial distortion, decentering distortion and thin prism distortion. Radial distortion commonly exceeds the other distortions by at least an order of magnitude,\cite{Hugemann2010a} and since the distortion is our images is minimal (maximum shift of about 15 pixels in a 4000x3000 pixel image) we only radial distortion in our correction.

Radial distortion is approximated by a polynomial of even powers:\cite{Hugemann2010a,Wang2008}
\begin{align}
\label{eq2a}
    \delta_{ur}(u,v) & = u(k_1r^2+k_2r^4+k_3r^6+\cdots) \\
    \label{eq2b}
    \delta_{vr}(u,v) & = v(k_1r^2+k_2r^4+k_3r^6+\cdots)
\end{align}
where $k_1,k_2,k_3$ are radial distortion coefficients; $r$ the distortion from a point $(u,v)$ to the center of radial distortion.\\
The first and second terms are predominant, so Equation \ref{eq2a} and \ref{eq2b} can be reduced to
\begin{align}
\label{eq3a}
    \delta_{ud} & = k_1 u_d r_d^2 + k_2 u_d r_d^4\\
    \label{eq3b}
    \delta_{vd} & = k_1 v_d r_d^2 + k_2 v_d r_d^4
\end{align}

We define $(x,y)$ as the coordinate system where $(x,y)=(0,0)$ is the top-left most corner of the image, as is common in image processing. We define $(u,v)$ as the coordinate system with the origin at the center of the image. $w,h$ are the width and height of the image in pixels, respectively. We assume that our distortion center (principal point) is located at the center of the image, so that
\begin{align}
    u &= x- \frac{w}{2}\\
    v &= y- \frac{h}{2}
\end{align}

\begin{figure}[!htbp]
\centering
  \includegraphics[width=1\linewidth]{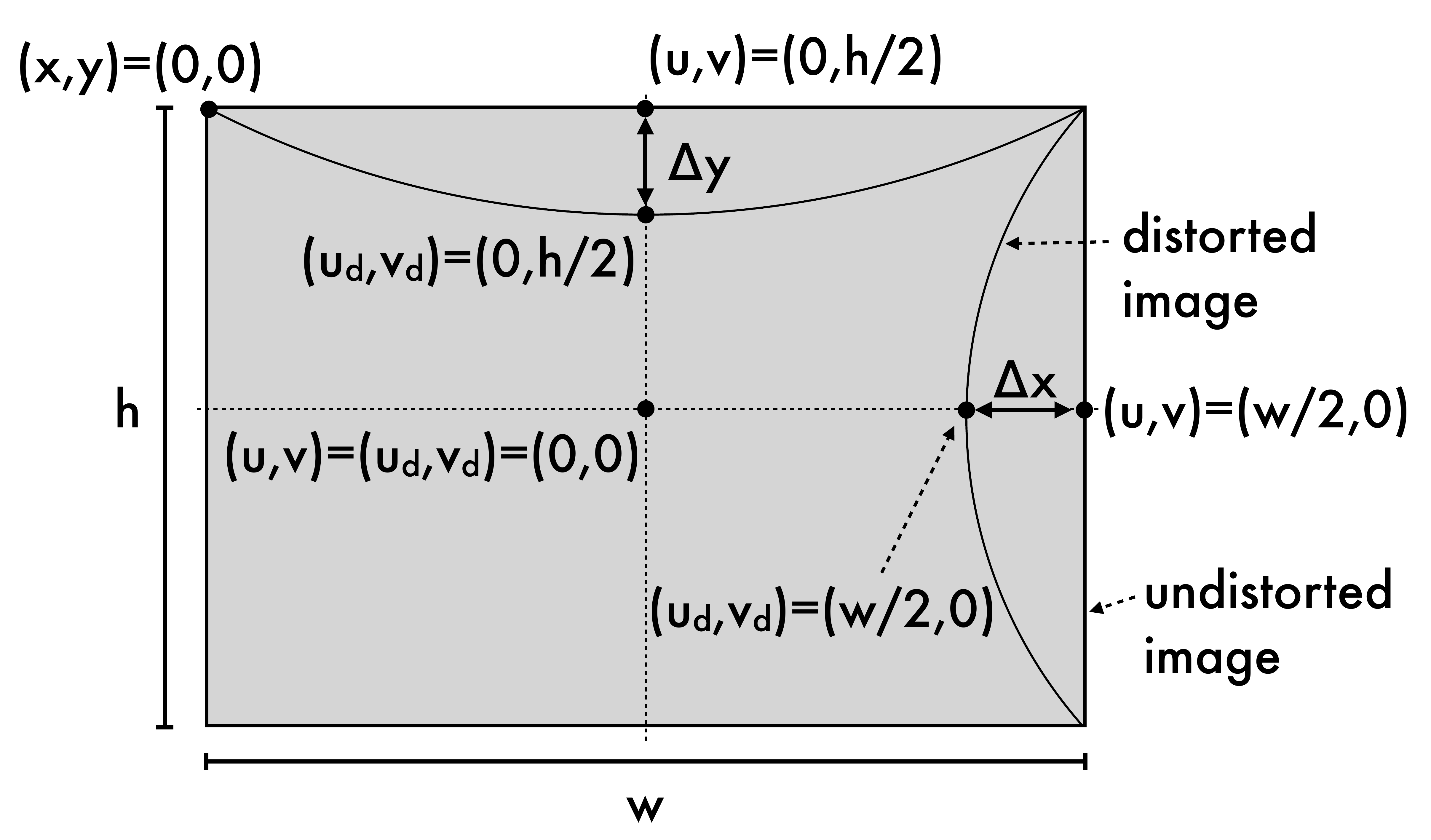}
\caption{Sketch of the image distortion math.}
\label{fig:imagedis}
\end{figure}

To determine the distortion coefficients $k_1$ and $k_2$, we measure the distortion using the straight electrodes in an experimental image at 2 points: $(u_d,v_d)=(0,h/2)$ and $(u_d,v_d)=(w/2,0)$. The distortion at these two points are $\Delta y = v_d-v$ and $\Delta x = u_d-u$, see image \ref{fig:imagedis}.

Substituting these 2 values in Equation \ref{eq3a} and \ref{eq3b} and solving these for $k_1$ and $k_2$, gives
\begin{align}
    k_1 &= \frac{8(-h^5 w^3 \Delta x +h^3 \Delta x (w+32\Delta x)+w^8 \Delta y}{h^3w^3(h^2w^3-w^5-32\Delta x)} \\
    k_2 &= \frac{-32w^3 \Delta y}{h^3(h^2w^3-w^5-32\Delta x)}
\end{align}

Using these in combination with Equation \ref{eq:1a}, \ref{eq:1b}, \ref{eq3a} and \ref{eq3b} allows us to remap all the drop $(x_d,y_d)$ locations to undistorted $(x,y)$ locations on a regular grid.

\section{Numerical simulation geometry}
\label{app:numerical_geometry}

\begin{figure}[!htbp]
\centering
  \includegraphics[width=1\linewidth]{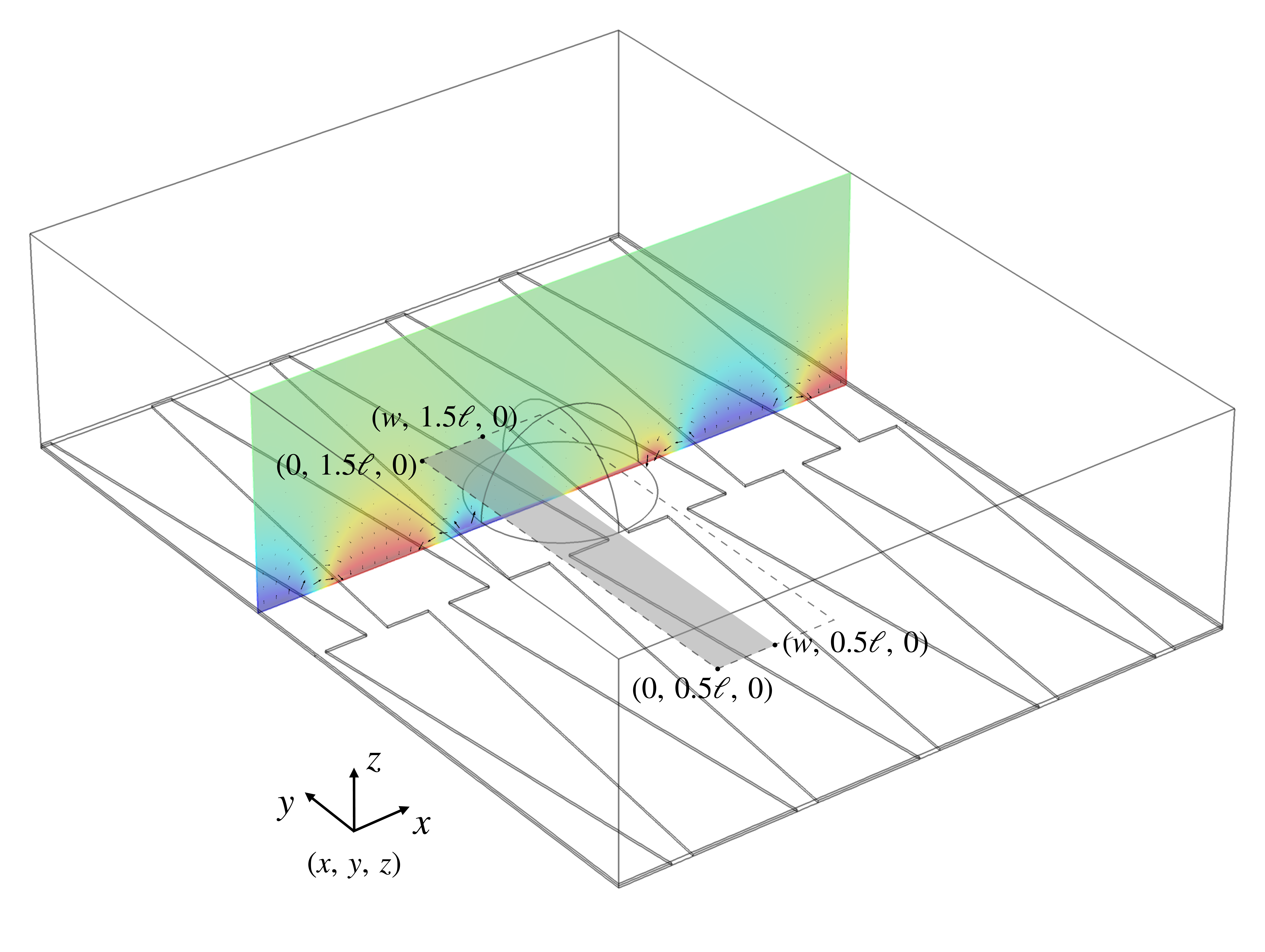}
\caption{Sketch of the computational domain for a medium-sized drop on electrodes with $\ell=1000\mu$m. The dashed rectangle indicates the surface unit cell, the gray shaded region indicates the limits of the $(x,y)$ drop center locations to create a full 2D electrostatic energy landscape. The color code in the cross section indicates the electrostatic potential and little arrows the electric field.}
\label{fig:numerical_geometry}
\end{figure}

\section{Horizontal transitions for other electrode lengths}
\label{app:horizontaltransitions_extra}
\begin{figure}[!htbp]
\centering
\includegraphics[width=1\linewidth]{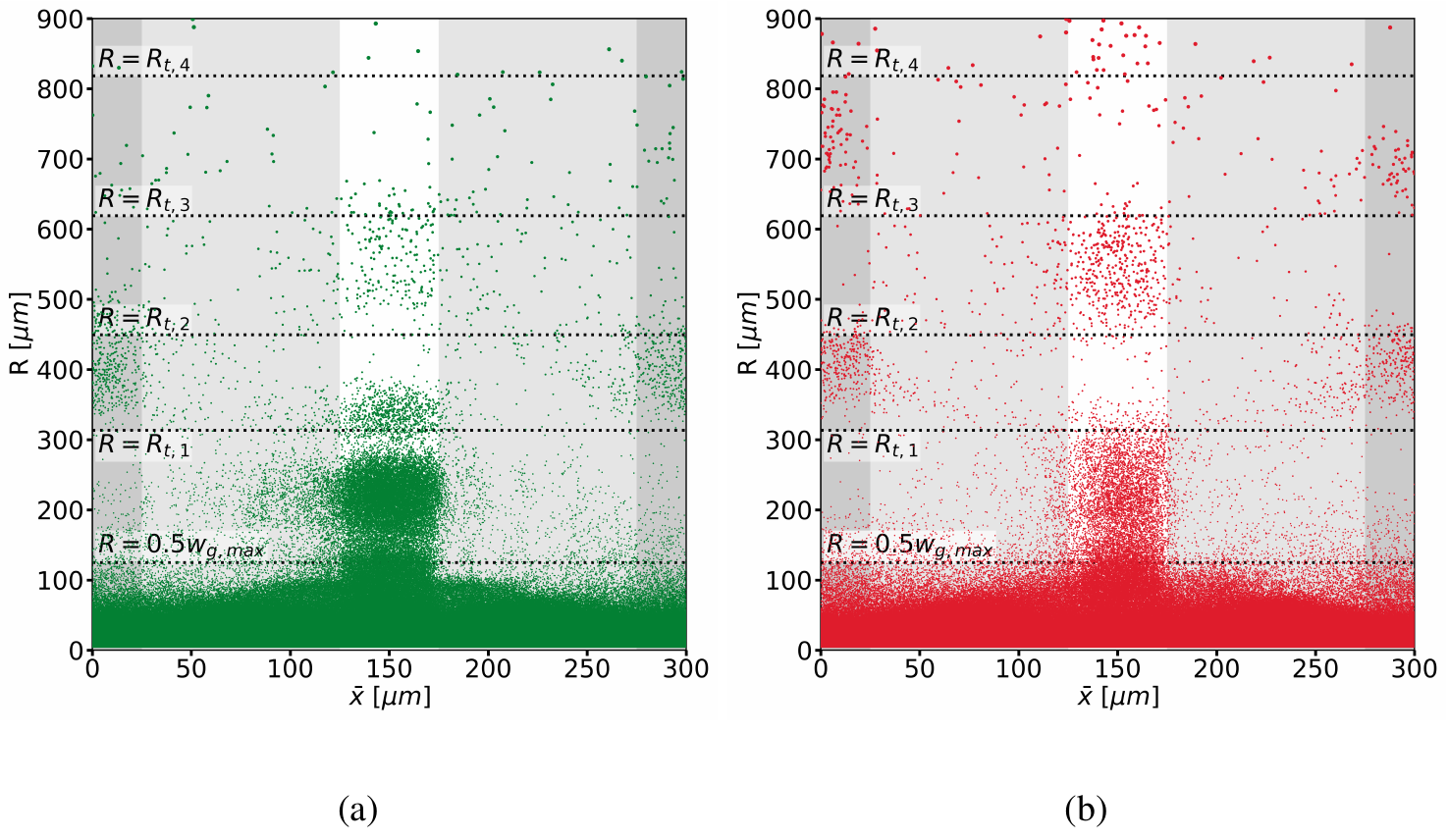}
\caption{Horizontal distribution of experimental drop centers (small dots) within a unit cell for variable drop radius $R$. Horizontal dotted lines indicate critical radii for first central alignment (0.5$w_{g,max}$) and subsequent transitions ($R_1...R_4$) from preferred alignment on gap center ($x=150\mu$m) and electrode center ($x=0$ or $300\mu$m). Bright and light shaded regions indicate the minimum and maximum gap widths $w_{g,min}$ and $w_{g,max}$. (a) data for $l=500\mu$m; (b) data for $l=3000\mu$m.}
\label{fig:horizontaltransitionsextra}
\end{figure}

see Figure \ref{fig:horizontaltransitionsextra}

\section{Surface degradation of Teflon surfaces}
\label{app:surfacedegradation}
\begin{figure}[!htbp]
\centering
  \includegraphics[width=1\linewidth]{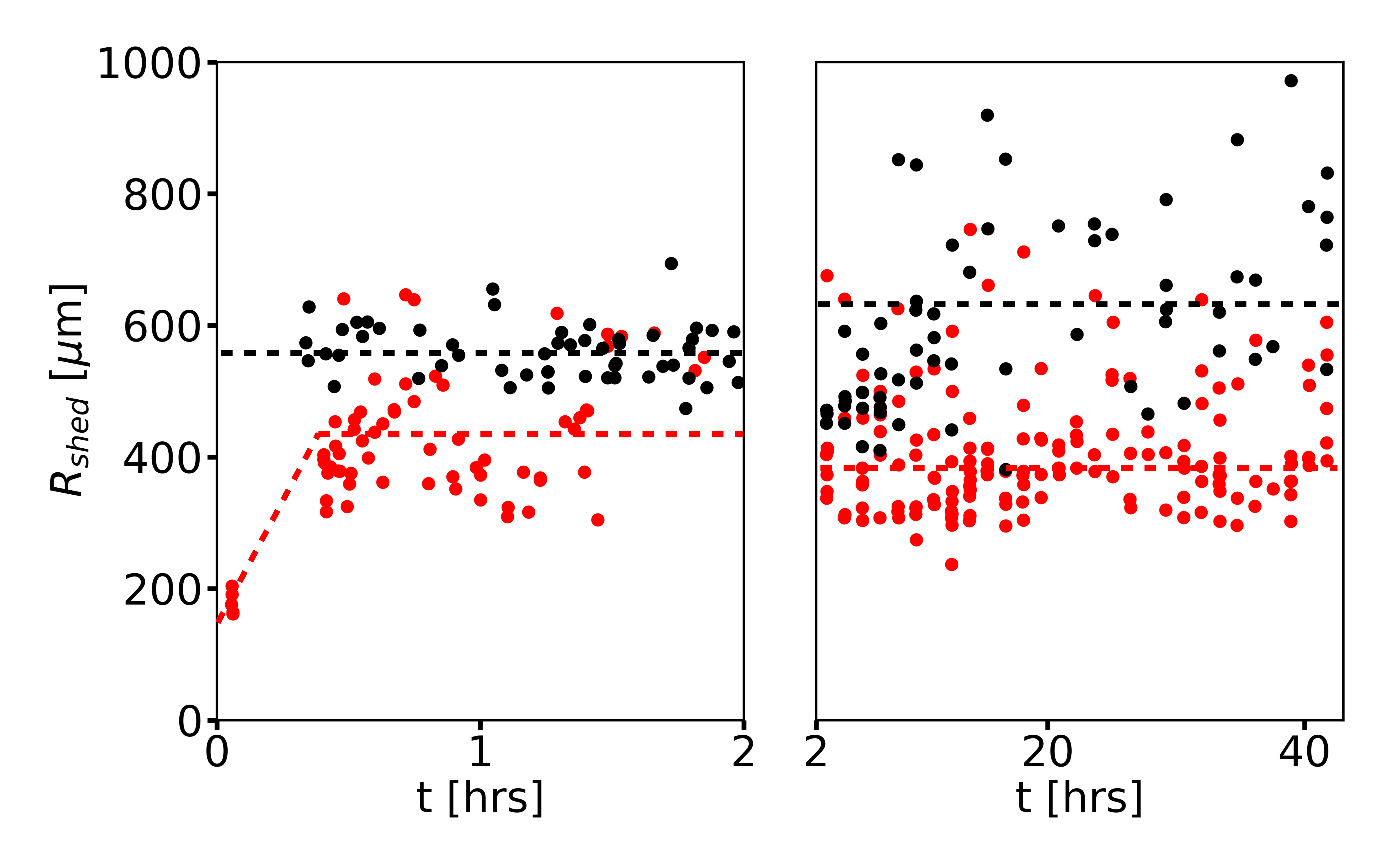}
\caption{Illustration of sample aging process: maximum radius of drops prior to shedding vs. time for Teflon AF samples for straight electrodes (gap width = electrode width = $200\mu$m). black: control; no EW, red = EW activated (150V; 1kHz). Dashed lines: guides to the eye.}
\label{fig:surfacedegradation}
\end{figure}

see Figure \ref{fig:surfacedegradation}
\\

\bibliographystyle{MSP}
\bibliography{library}
\end{document}